\documentstyle[aps,epsf,twocolumn]{revtex}
\begin{document}
\twocolumn[\hsize\textwidth\columnwidth\hsize\csname
@twocolumnfalse\endcsname

\title{Quantum spin hydrodynamics and a new spin-current mode in ferromagnetic
metals}
\author{Kevin S. Bedell}
\address{Department of Physics, Boston College, Chestnut Hill, MA 02167}
\author{Krastan B. Blagoev}
\address{Department of Radiology, Brigham and Women's
Hospital and Harvard Medical School, 75 Francis St., Boston MA 02115}
\date{\today}
\maketitle
\begin{abstract}

We derive the quantum spin hydrodynamic equations in a ferromagnetic metal.
From these equations we show the existence of a new massive spin-current mode.
This mode can be observed in neutron scattering experiments and we discuss
the difficulties in seeing it. At the end we discuss the existence of this mode
in localized ferromagnets.

\end{abstract}
]
Propagating collective modes in condensed matter systems are often a
consequence of spontaneously broken symmetry. In ferromagnetic systems, the
spin rotational symmetry is spontaneously broken leading to the existence of
spin waves. They were first observed in iron using neutron scattering
experiments in the nineteen fifties\cite{Lowde1956}. Theoretically spin
waves were predicted\cite{Bloch1930,Slater1930} in lattice models with a
ferromagnetic ground state as well as in ferromagnetic Fermi 
liquids\cite{Abrikosov and Dzyaloshinskii1959}. In paramagnetic metals, spin waves
propagate in an external magnetic
field\cite{Silin1957}.
Not all propagating modes are a consequence of spontaneously broken
symmetry and an example is zero sound in liquid helium\cite{Landau1956}.

In this paper we will demonstrate the existence of a new collective spin mode in
ferromagnetic metals. We start from a quasi-classical kinetic equation, the
Landau kinetic equation, for the momentum distribution function in a ferromagnetic
metal. From this we can derive the dynamic equations for the evolution of 
the spin density and spin current. In the low temperature limit the spin
dynamics is collisionless and dominated by spin precession, and we will call
this the quantum spin hydrodynamic (QSH) regime. At higher temperature this dynamics
crosses over to the classical spin hydrodynamics (CSH) of Halperin and
Hohenberg\cite{Halperin and Hohenberg1969}. In the QSH regime
we show that a new massive mode
exists along with the usual magnon. While gaped, it exists outside of the
particle-hole, Stoner, continuum and therefore it can propagate. From a macroscopic
point of view, this mode is induced by the collective oscillations of the spin current
and it should be observable in neutron scattering experiments.

In the long wave length limit one can derive the existence of the spin waves
from the conservation law\cite{Halperin and Hohenberg1969} for the
magnetization in the CSH limit. 
However, the spin current is not a conserved quantity and
therefore one cannot derive propagating modes related to the fluctuations of
the spin current from the conservation laws. Here the Landau kinetic equation
first used by Abrikosov and 
Dzyaloshinksii\cite{Abrikosov and Dzyaloshinskii1959} for weak ferromagnetic metals
is used to derive the full QSH theory.

We consider a three dimensional, weak itinerant ferromagnet below its Curie
temperature. To avoid complications we ignore the effects related to the
lattice. The phenomena which we will describe is independent of the lattice,
although lattice effects can add more features to the ones that we study. We
assume that the itinerant ferromagnet is well described by ferromagnetic
Fermi liquid theory\cite{Abrikosov and Dzyaloshinskii1959,Abrikosov1986}. 
The kinetic equations for the
magnetization and the spin current can be derived from the equation for the
Fourier transform of the Wigner distribution function 
$n_{\vec{p}}^{\alpha \beta }(\vec{r},t)$\cite{Silin1957} 
\begin{equation}
\frac{\partial n_{\vec{p}}^{\alpha \beta }(\vec{r},t)}{\partial t}-
\frac 1{i\hbar }F_{\vec{p}}^{\alpha \beta }[n;\epsilon ]=I_Q[n_{\vec{p}^{\prime }}],
\end{equation}
where 
\begin{eqnarray}
F_{\vec{p}}^{\alpha \beta }[n;\epsilon ] &=&\sum_\gamma \int Dp^{\prime
}Dqd^3r^{\prime }d^3\rho e^{i\left[ (\vec{p}-\vec{p}^{\prime })\cdot 
\vec{\rho}+\vec{q}\cdot (\vec{r}-\vec{r}^{\prime })\right] }  \nonumber \\
&&\times (\epsilon _{\vec{p}^{\prime }+\hbar \vec{q}/2}^{\alpha \gamma }
(\vec{r}^{\prime }+\frac{\hbar \vec{\rho}}2,t)n_{\vec{p}^{\prime }}^{\gamma
\beta }(\vec{r}^{\prime },t)  \nonumber \\
&&-n_{\vec{p}^{\prime }}^{\alpha \gamma }(\vec{r}^{\prime },t)
\epsilon _{\vec{p}^{\prime }-\hbar \vec{q}/2}^{\gamma \beta }(\vec{r}^{\prime }-
\frac{\hbar \vec{\rho}}2,t)),
\end{eqnarray}
\begin{equation}
n_{\vec{p}}^{\alpha \beta }(\vec{r},t)=
\int Dqe^{i\vec{q}\cdot \vec{r}}\left\langle a_{\vec{p}-\vec{q}/2,\alpha }^{\dag }
a_{\vec{p}+\vec{q}/2,\beta }\right\rangle ,
\end{equation}
with $a_{\vec{p}+\vec{q}/2,\beta }$ the single fermion operator for a
particle with spin $\beta $. Here $Dq=\frac{d^3q}{(2\pi \hbar )^3}$ and
$\epsilon _{\vec{p}}^{\alpha \beta }(\vec{r},t)$ is the effective
quasiparticle Hamiltonian. Also $I_Q$ is the quantum collision integral which
contains contributions from two body collisions and from the rate of change due
to the spin precession\cite{Baym and Pethick1991}. 

Following Ref.\cite{Baym and Pethick1991} 
we expand all single body operators in the basis set spanned by the
Pauli matrices 
\begin{equation}
\epsilon _{\vec{p}}^{\alpha \beta }(\vec{r},t)=
\epsilon _{\vec{p}}\delta^{\alpha \beta }(\vec{r},t)+
\vec{h}_{\vec{p}}(\vec{r},t)\cdot \vec{\tau}^{\alpha \beta },
\end{equation}
\begin{equation}
n_{\vec{p}}^{\alpha \beta }(\vec{r},t)=
n_{\vec{p}}\delta ^{\alpha \beta }(\vec{r},t)+
\vec{\sigma}_{\vec{p}}(\vec{r},t)\cdot \vec{\tau}^{\alpha \beta }
\end{equation}
at every point $(\vec{r},t)$ of space and time with $\vec{\sigma}_{\vec{p}}$ the
magnetization and 
\begin{equation}
\vec{h}_{\vec{p}}=-\gamma \frac \hbar 2\vec{{\cal {H}}}+2\int Dp^{\prime }
f_{\vec{p}\vec{p}^{\prime }}^a\vec{\sigma}_{p^{\prime }}
\end{equation}
the effective magnetic field which is a sum of an external magnetic field ${\cal \vec{H}}$ 
and an internal magnetic field generated by the quasiparticle
interactions $f^a_{\vec{p}\vec{p}^{\prime }}$. We expand the quasiparticle interactions
in spherical harmonics and separate the spin symmetric ($f_l^s$)
and spin antisymmetric ($f_l^a$) parts. The result is
\begin{equation}
N(0)f_{\vec{p}\vec{p}^{\prime }}^{{\bf \tau \tau }^{\prime
}}=\sum_l(F_l^s+F_l^a\vec{\tau}\cdot \vec{\tau }^{\prime })P_l( 
\hat{p}\cdot \hat{p}^{\prime }),
\end{equation}
with
\begin{equation}
f_{\vec{p}\vec{p}^{\prime }}^{\tau \tau ^{\prime }}
\approx f_{\vec{p}\vec{p}^{\prime}}^s+
f_{\vec{p}\vec{p}^{\prime }}^a\vec{\tau}\cdot \vec{\tau}^{\prime }.
\end{equation}
Here we treat the quasiparticle interaction as spin rotation
invariant. This is possible in the case of small magnetizations and external
magnetic fields as explained in\cite{Blagoev et al.1999}.
Here $N(0)$ is the average density of states, averaged
over the two Fermi surfaces. We must note that the
condition that the ground state of the metal is ferromagnetic means that
$N(0)f_0^a=F_0^a<-1$. 

For an inhomogeneous system the
spin quantization axis is position dependent and therefore the spin
density 
\begin{equation}
\vec{\sigma}(\vec{r},t)=2\int Dp\vec{\sigma}_{\vec{p}}(\vec{r},t)
\end{equation}
is position dependent. From the kinetic equation one can obtain
the continuity equation for the spin density which reduces to the usual
spin conservation law 
\begin{equation}
\frac{\partial \vec{\sigma}(\vec{r},t)}{\partial t}+\frac \partial {\partial
x_i}j_{\vec{\sigma},i}(\vec{r},t)=
\gamma \vec{\sigma}(\vec{r},t)\times {\cal \vec{H}},
\end{equation}
where the spin current is 
\begin{equation}
j_{\vec{\sigma},i}(\vec{r},t)=
2\int Dp\left[ \frac{\partial 
\epsilon _{\vec{p}}}{\partial p_i}\vec{\sigma}_{\vec{p}}(\vec{r},t)+
\frac{\partial \vec{h}_{\vec{p}}}{\partial p_i}n_{\vec{p}}(\vec{r},t)\right]
\end{equation}
and a summation over repeated indices is used. 
The precession term in the $r.h.s.$ of Eq.(10) follows from the
non-commutativity of the particle density operator and the quasiparticle 
Hamiltonian. Assuming that the magnetization is small and that we are
close to equilibrium we can linearize the spin current which leads to
\begin{equation}
j_{\vec{\sigma},i}(\vec{r},t)=2\int Dp v_{\vec{p}_i}\vec{\sigma}_{\vec{p}}
\left(1+\frac{F_1^a}{3}\right),
\end{equation}
because in the Fermi liquid only momenta in the vicinity of the Fermi surface
contributes to the magnetization. 
Here $\vec{v}_p=\vec{\nabla}_p\epsilon_p^0$, $n_p^0$ is the equilibrium 
distribution function, and $F_1^a$ is a Fermi liquid parameter.

Starting from the kinetic equation for the spin 
density distribution function\cite{Baym and Pethick1991}
\begin{eqnarray}
\frac{\partial \vec{\sigma}_{\vec{p}}}{\partial t}+
v_{\vec{p}_i}\frac{\partial}{\partial r_i}
\left(\vec{\sigma}_{\vec{p}}-
\frac{\partial n_p^0}{\partial \epsilon_p}\vec{h}_{\vec{p}}\right)&=& \nonumber \\
-\frac{2}{\hbar}\vec{\sigma}_{\vec{p}}\times\vec{h}_{\vec{p}}+
\left(\frac{\partial\vec{\sigma}_{\vec{p}}}{\partial t}\right)_{coll.}
\end{eqnarray}
then multiplying by the quasiparticle velocity and integrating
over the quasiparticle momentum the linearized equation for 
the spin current takes the form
\begin{equation}
\frac \partial {\partial t}j_{\vec{\sigma},i}(\vec{r},t)+
\frac \partial {\partial x_m}\Pi _{\vec{\sigma}im}=
\left[ \frac \partial {\partial t}j_{\vec{\sigma},i}\right] _{prec.}+
\left[ \frac \partial {\partial t}j_{\vec{\sigma},i}\right] _{coll.},
\end{equation}
where the spin stress tensor is 
\begin{equation}
\Pi _{\vec{\sigma}im}=2\left( 1+\frac{F_1^a}3\right) \int
Dpv_{p_i}v_{p_m}\left( \vec{\sigma}_{\vec{p}}-
\frac{\partial n_{\vec{p}}^0}{\partial \epsilon _p}\vec{h}_{\vec{p}}\right).
\end{equation}

Next we apply a small, transverse (to the external field and therefore to the
magnetization) magnetic field
$\delta\vec{{\cal H}}=\delta {\cal H}_x\hat{e}_x+\delta {\cal H}_y\hat{e}_y$ 
which induces a change in the magnetization
$\delta \vec{\sigma}_{\vec{p}}=
\delta \sigma _{\vec{p}x}\hat{e}_x+\delta \sigma _{\vec{p}y}\hat{e}_y$, where 
$\hat{e}_i$ are the unit vectors and 
$\delta \vec{\sigma}_{\vec{p}}=\vec{\sigma}_{\vec{p}}-\vec{\sigma}_{\vec{p}}^0$. 
To make the expansion in spherical harmonics
more transparent we define
\begin{equation} 
\delta \sigma^{\pm }_{\vec{p}}=
-\frac{\partial n_{\vec{p}}^0}{\partial\epsilon_p^0}\nu_{\vec{p}}^{\pm}
\end{equation}
and
\begin{equation}
\sigma^0_{\vec{p}}=-\frac{\partial n_{\vec{p}}^0}{\partial\epsilon_p^0}\frac{m}{N(0)},
\end{equation}
where $\delta \sigma^{\pm }_{\vec{p}}=\delta\sigma _{\vec{p}x}\pm i\delta \sigma _{\vec{p}y}$, 
$\sigma^0_{\vec{p}}$
is the magnitude of the magnetization in the absence of the perturbing field, 
and $m$ is the equilibrium magnetization. We expand in spherical harmonics the quantities
\begin{equation}
\nu_{\vec{p}}^{\pm}=\sum_l\nu_l^{\pm}P_l(\hat{p}\cdot \hat{k})
\end{equation}
and substitute the corresponding expressions in Eq.(13). Here $\hat{k}$ is the unit vector
in the direction of the spin quantization axis. Now Eq.(13) involves terms
grouped according to the value of the integer number $l$. Next we neglect the terms with $l>1$.
Although the term $l=2$ is very important as we discuss later, the derivation which
we present below is clearer and contains all the steps needed for the derivation
of the dispersion including higher spherical harmonics.
Then using the expansion in spherical harmonics, Eq.(18), and in the presence of 
an external magnetic field, Eq.(14) reduces to
\begin{eqnarray}
&&\frac \partial {\partial t}\vec{j}_{\vec{\sigma},i}+
\left( 1+\frac{F_1^a}3\right) \left| 1+F_0^a\right| \frac{v_F^2}3
\frac \partial {\partial x_i}\vec{\sigma}(\vec{r},t)  \nonumber \\
&=&\gamma \vec{j}_{\vec{\sigma},i}\times {\cal H-}\frac 2\hbar 
\left( f_0^a-\frac{f_1^a}3\right) 
\vec{j}_{\vec{\sigma},i}\times \vec{\sigma}(\vec{r},t)-\left( 1+
\frac{F_1^a}3\right) \frac{\vec{j}_{\vec{\sigma},i}}{\tau _D},  \nonumber \\
&&
\end{eqnarray}
Here we have used a relaxation time approximation for the
collision contribution to the spin current. 

Before we examine the full spin dynamics coming from Eqs.(10) and (19) it is
useful to contrast this with the behavior of a paramagnetic metal. In a
paramagnetic Fermi liquid the first two terms on the $r.h.s.$ of Eq.(19) are
proportional to $\omega_L=2\gamma\cal H$, where $\cal H$ is the external
magnetic field. At high temperature the CSH regime is realized and at low
temperature the system crosses over to the precession dominated regime.
The cross over scale is set by $\omega_L\tau_D\approx1$. This behavior is
similar to that in paramagnetic 
metals\cite{Baym and Pethick1991,Leggett and Rice1968}. 
In the ferromagnetic
metal the cross over scale is nearly independent of the external magnetic field.
To see this we first note that at high temperature, Eq.(19) in the uniform
case reduces to
\begin{equation}
\frac \partial {\partial t}\vec{j}_{\vec{\sigma},i}=
-\left( 1+\frac{F_1^a}3\right) \frac{\vec{j}_{\vec{\sigma},i}}{\tau _D}
\end{equation}
which means that $\vec{j}_{\vec{\sigma},i}$ exponentially decays and its
explicit time dependence can be ignored.
In this limit, which is called the classical
spin hydrodynamic limit with 
$\frac{\partial \vec{j}}{\partial t}=0$ the 
equations reduce to those used by Halperin and 
Hohenberg\cite{Halperin and Hohenberg1969} and it is easy to derive the dispersion,
$\omega \sim q^2$ for the spin waves\cite{Abrikosov and Dzyaloshinskii1959}.
The quadratic dependence for the Goldstone mode follows from the fact
that the order parameter $m$ is a conserved charge of the spin rotation
algebra. As the system crosses over to low temperature it enters into the QSH 
regime in which the spin current changes in time due to the precession around
the effective field with $\omega_1^+=2m(f_1^a/3-f_0^a)$. For zero external field
the cross over from QSH to CSH occurs when 
$\omega_1^+\tau_D\approx1$. The range of temperatures where $\omega_1^+\tau_D>>1$
is what we have referred to as the QSH regime. In this QSH regime we can not
ignore the $\frac \partial {\partial t}\vec{j}_{\vec{\sigma},i}$ term in Eq.(19).
This as we will see is the source of the new spin-current mode.

There is another way to see the difference between spin waves in
a ferromagnetic and a paramagnetic metal.
Consider a ferromagnetic metal in an external
magnetic field. The magnetization of the metal will align
with the external field. If we apply a small pulse in a direction not collinear 
with the external magnetic field, 
the magnetization will start to precess around the
external field. Next we reduce to zero the external magnetic field.
In a paramagnetic metal the polarization will also go to zero and the
spin wave will become damped. However,
in a ferromagnetic metal the magnetization will continue to precess
about the same axis, although no external magnetic field is present
to define that axis. Now one can imagine looking at the system
from a frame of reference rotating with the magnetization. In this
frame of reference there will be no magnetization precession, but
there will be a precession of the spin current, because the spin current
is not a conserved quantity and as seen from Eqn.(19) it precesses
with the frequency $\omega_1^+$. 

It is instructive to look first at the modes of the system in the simpler case
when $F_l^a=0$, for $l>1$ and the external field is ${\cal H}=0$. In that case the
equations simplify to
\begin{equation}
\frac{\partial \delta \vec{\sigma}(\vec{r},t)}{\partial t}+
\frac \partial {\partial x_i}j_{\vec{\sigma},i}(\vec{r},t)=0,
\end{equation}
\begin{eqnarray}
\frac \partial {\partial t}\vec{j}_{\vec{\sigma},i}+
c_s^2\frac \partial {\partial x_i}\delta \vec{\sigma}
&&=-\frac 2\hbar (f_0^a-\frac{f_1^a}{3})\vec{j}_{\vec{\sigma},i}\times \left( \vec{\sigma}^0+
\delta \vec{\sigma}\right) \nonumber \\
&&-(1+\frac{F_1^a}{3})\frac{\vec{j}_{\vec{\sigma},i}}{\tau _D},
\end{eqnarray}
where $c_s^2=\left| 1+F_0^a\right| (1+\frac{F_1^a}3)\frac{v_F^2}3$ is the
spin wave velocity, with $v_F$ the Fermi velocity. Here we have introduced 
$\delta \vec{\sigma}=\vec{\sigma}(\vec{r},t)-\vec{\sigma}^0$, 
with $\vec{\sigma}^0=m\hat{k}$, where $m$ is the magnetization.
If the system is in the QSH regime
and $\delta \vec{\sigma}\ll \vec{\sigma}^0$ one can write for
the spin current the following hydrodynamic equation 
\begin{equation}
\frac \partial {\partial t}\vec{j}_{\vec{\sigma},i}+
c_s^2\frac \partial {\partial x_i}\delta \vec{\sigma}=
{\cal -}\frac 2\hbar (f_0^a-\frac{f_1^a}{3})\vec{j}_{\vec{\sigma},i}\times \vec{\sigma}^0.
\end{equation}
Differentiating this equation with respect to $x_i$ and using the equation
for the spin density we obtain 
\begin{equation}
\frac{\partial ^2}{\partial t^2}\delta \vec{\sigma}+
c_s^2\frac{\partial ^2}{\partial x_i^2}\delta \vec{\sigma}+
\frac 2\hbar (f_0^a-\frac{f_1^a}{3})
\frac{\partial \delta \vec{\sigma}}{\partial t}\times \vec{\sigma}^0=0.
\end{equation}
Below we will set $\hbar =1$. Applying again a transverse magnetic field
as above Eq.(16) we can write 
\begin{equation}
\frac{\partial ^2\delta \sigma^{+}}{\partial t^2}+
c_s^2\frac{\partial^2\delta \sigma^{+}}{\partial x_i^2}-
2im(f_0^a-\frac{f_1^a}{3})\frac{\partial \delta \sigma^{+}}{\partial t}=0.
\end{equation}
Assuming that the solution is of the form 
\begin{equation}
\delta \sigma^{+}=m^{+}e^{i(\vec{q}\cdot \vec{r}-\omega t)}
\end{equation}
with $\vec{q}=q\hat{e}_x$ we obtain the dispersion equation 
\begin{equation}
\omega ^2+c_s^2q^2+2m(f_0^a-\frac{f_1^a}{3})\omega =0.
\end{equation}
The two modes have the dispersion 
\begin{equation}
\omega _l=\left| m(f_0^a-\frac{f_1^a}{3})\right| 
\left( 1\pm \sqrt{1-\frac{c_s^2q^2}
{\left|m(f_0^a-\frac{f_1^a}{3})\right| ^2}}\right) .
\end{equation}
where the $l=0(+)$ solution corresponds to the spin waves 
and the $l=1(-)$ solution corresponds to a new collective spin mode. 
For small $q$ we have the usual Goldstone mode 
\begin{equation}
\omega _0(q)=\frac{c_s^2}{\omega_1^{+}}q^2
\end{equation}
and the new mode has the dispersion
\begin{equation}
\omega _1(q)=\omega _1^{+}-\frac{c_s^2}{\omega _1^{+}}q^2.
\end{equation}
Under close examination one 
finds that to the $q^2$ term there is a
contribution from the $l=2$ spherical harmonics in Eq.(18). 
Taking that into account
gives the full expression for this mode 
\begin{equation}
\omega _1(q)=\omega _1^{+}-\left[ \frac{c_s^2}{\omega _1^{+}}+
\frac{4N(0)v_F^2}{30m}\left(\frac{3}{F_1^a}+1\right)\right] q^2.
\end{equation}
The above equation is valid for $qv_F<<m|f^a_1|$. In the limit $f_1^a\rightarrow0$
the mode merges with the Stoner continuum and is Landau damped.
For $f_1^a\neq0$ the mode propagates with a dispersion that depends
on the sign of $f_1^a$.

It is useful to look at the contribution of this mode to the $f$-sum rule.
It is known that the spin waves do not satisfy this sum rule. However the spin 
waves plus the new mode exhaust the $f$-sum rule and
the spectral weight is shifted from the Stoner excitations to this new mode.
To see this we first introduce the dynamic structure function 
\begin{equation}
S^{\tau\tau^{\prime}}(\vec{q},\omega )=-\frac 1\pi 
\mathop{\rm Im}
\chi ^{\tau\tau^{\prime}}(\vec{q},\omega )\theta (\omega ),
\end{equation}
where $\chi (\vec{q},\omega )$ is the dynamic spin susceptibility 
obtained from the kinetic equation by noting that 
$\chi=\frac{\delta\sigma}{\delta{\cal H}}$. Here $\theta(\omega )$ 
is the step function. 
From the expression for the dynamic structure function, calculated
from the kinetic equation Eq.(13), we can derive the f-sum rule for
the transverse dynamic spin response function, $S^{+-}(\vec{q},\omega)$.
This is given by 
\begin{equation}
\int_0^\infty d\omega \omega S^{+-}(\vec{q},\omega )=
(1+\frac{F_1^a}3)\frac{nq^2}{2m^{*}},
\end{equation}
where $n$ is the total density and $m^*$ is the effective mass.
This has the same form as the one obtained for the spin dynamic
structure function in the paramagnetic 
phase\cite{Baym and Pethick1991,Leggett1966}.

This sum rule is not exhausted by the Goldstone mode alone. If
we now include the new mode at zero temperature we have
\begin{equation}
S^{+-}(\vec{q},\omega )=\alpha_{\vec q}^{+-}\delta (\omega -\omega _0(q))
+\beta_{\vec q}^{+-}\delta (\omega -\omega_1(q))
\end{equation}
and this exhausts the $f$-sum rule, Eq.(33). This is significant, because
the spectral weight of the Stoner excitations has been transfered to
the two modes, leaving 
\begin{equation}
\int_0^\infty d\omega \omega S_{Stoner}^{+-}(\vec{q},\omega )\propto q^4.
\end{equation}
This might be a plausible explanation for the difficulty of the
direct observation of the Stoner excitations in a neutron scattering
experiments\cite{Mook1988} in Fe and Ni since the 
oscillator strength of these single excitations has been reduced.

The observation of the new mode described in this paper is difficult because 
it has a small spectral weight at small $q$. This follows from the
$q^2$ dependence of $\beta_{\vec q}^{+-}$ term in Eq.(34). Another
obstacle in observing the mode is that it originates from
the oscillations of the spin current, which is not a conserved
quantity and therefore it is easily damped.
 
In conclusion, in this paper we described a new collective mode in weak 
ferromagnetic metals in the quantum spin hydrodynamic regime.
While our calculations are for small moment
itinerant ferromagnets this
mode will exist in other ferromagnetic metals in which the moment
is not small. In that case the calculations are 
significantly more involved. 
This new mode should exist as well in local moment ferromagnets. 
Mathematically this can be seen
if one relaxes the condition
used by Halperin and Hohenberg\cite{Halperin and Hohenberg1969}
that the partial time derivative of the spin current is zero.
This as well as a calculation of this mode at finite temperature
we leave for a future publication.

We would like to thank A. Balatsky, A. Bishop, R. Cowley, 
P.B. Littlewood, R. McQeeney, J.L. Smith, and S. Trugman,
for the fruitful discussions. 
P. Petkova participated at the
early developments of this paper and we would like to
thank her for the discussions during that period. 
K.S. Bedell would like to thank the Aspen Center for 
Theoretical Physics where some of the ideas developed in 
this paper were born.
This work was supported by the DOE Grant DEFG0297ER45636.

\end{document}